# Field-free superconducting diode effect in layered superconductor FeSe

Utane Nagata[1], Motomi Aoki[1,2], Akito Daido[3], Shigeru Kasahara[4], Yuichi Kasahara[3], Ryo Ohshima[1,2], Yuichiro Ando[1,2], Youichi Yanase[2,3], Yuji Matsuda[3] and Masashi Shiraishi[1,2,†]

[1]Department of Electronic Science and Engineering, Kyoto University, Kyoto, Kyoto 615-8510, Japan

[2]Center for Spintronics Research Network, Institute for Chemical Research, Kyoto University, Uji, Kyoto 611-0011, Japan

[3]Department of Physics, Graduate School of Science, Kyoto University, Kyoto 606-8502, Japan

[4] Research Institute for Interdisciplinary Science, Okayama University, Okayama, 700-8530, Japan

**Corresponding authors**

Masashi Shiraishi

E-mail: shiraishi.masashi.4w@kyoto-u.ac.jp

**The superconducting diode effect (SDE), where zero-resistance states appear nonreciprocally during current injection, is receiving tremendous interest in both fundamental and applied physics because the SDE is a novel manifestation of symmetry breaking and enables the creation of a novel diode. In particular, magnetic-field-free SDEs have been extensively investigated because of their potential to serve as building blocks for superconducting circuit technology. In this letter, we report the field-free SDE in a layered superconductor, FeSe. Its underlying physics is clarified by systematic controlled experiments to be an interplay of a large thermoelectric response and geometrical asymmetry in FeSe. Our findings can pave a new avenue for the construction of novel material and device platforms utilizing SDEs.**

Nonreciprocal transport is a family of rectification effects in which the amplitudes of the electric current flowing in the opposite directions are not the same; specifically, a nonlinear response beyond Onsager's reciprocity occurs [1–3]. Nonreciprocal transport manifests itself as a consequence of breaking of the spatial inversion and time reversal symmetries, and this novel function in charge transport has attracted the attention of many researchers [4–8] since the first experimental demonstration of nonreciprocal charge transport in a Si metal-on-semiconductor field effect transistor [3]. The superconducting diode effect (SDE), the manifestation of nonreciprocity in the critical current, is categorized as a family of nonreciprocal charge transport [9]. In the SDE, the spatial inversion and time reversal symmetries are simultaneously broken by the so-called tricolor stacking of Nb/V/Ta and an external in-plane magnetic field, respectively; both of these are indispensable for nonreciprocal transport. In theory, the broken spatial inversion and the time reversal symmetries allow the Cooper pairs in the system to acquire finite momentum; this process is the inherent mechanism of the SDE [10–12] and signifies the importance of the symmetry breaking combination. In subsequent studies, many material platforms have been reported for the appearance of the SDE [13–15], and a variety of other mechanisms, such as asymmetric tunnelling at the Josephson junction [16–19], magnetochiral anisotropy (MCA) at the surface state of topological superconductors [20], and Josephson vortices in multilayered superconducting nanowires [21], have been proposed. Furthermore, numerous research has been performed to explain the detailed underlying physics from theoretical perspectives [22–24].

Importantly, although the SDE platforms reported thus far commonly possess structural asymmetry (interfaces, junctions, and crystal structures with a broken inversion symmetry), a magnetic field to break time reversal symmetry is not always required. Indeed, the field-free Josephson diode effect (JDE) [16], where an external magnetic field to break the time reversal symmetry is not applied, takes place in some material systems with inversion symmetry breaking, and its underlying physics is still elusive. Furthermore, a recent systematic investigation of the SDE [25] revealed that macroscopic asymmetries such as asymmetric channel geometry, i.e. extrinsic mechanisms, can allow the nonreciprocity of the critical current, which requests revisiting the conventional SDE results. Thus, a quest of further precise understandings of the physics behind the SDE that is currently somewhat intricately intertwined is strongly awaited and significant.

In this letter, we report the observation of a field-free SDE in a layered superconductor, FeSe, with a symmetric crystal structure. Although the spatial inversion symmetry in the crystal and time reversal symmetry are intrinsically conserved in FeSe, the salient SDE is surprisingly observed. We successfully corroborated that the field-free SDE in this work was attributed to the interplay of a sample-geometry-induced thermal gradient and thermoelectric effect in FeSe, which is clearly supported by the experimental finding of sign inversion of diode efficiencies by reversing the total heat flow in FeSe using an external heating wire equipped to the SDE device. Our study elucidates a novel physical aspect of the SDE, the thermoelectric contribution, and enables the expansion of material platforms for the SDE. Furthermore, our study enables delving further into the field-free SDEs by requesting to revisit the sample geometries and thermoelectric characteristics of the materials introduced thus far.

Figure 1(a) shows a schematic of the SDE appearing in an FeSe flake. FeSe is a family of iron-chalcogenide superconductors that possess layered and spatially symmetric crystal structures (see Fig. 1(b)) [26]. FeSe was exfoliated and transferred onto a thermally oxidized Si substrate. The thickness of the FeSe flake was approximately 114 nm. Since the superconducting transition is suppressed when FeSe is thinner than 20-30 nm, which is almost consistent with the result shown in ref. [27], we set the thickness of the FeSe to be greater than 30 nm. Rectangular Au (100 nm) electrodes were fabricated on the FeSe flakes by using electron-beam (EB) lithography and EB deposition following $Ar^+$ ion milling in a load-lock chamber enabling Ohmic contact with the FeSe. Nonmagnetic Ti (3 nm)/Au (150 nm) pads connected to the Au electrodes on the FeSe were fabricated using EB lithography and EB deposition. Notably, neither the Josephson junction nor the stacking structures that break inversion symmetry existed in the device, and ferromagnetic electrodes to break the time reversal symmetry were not exploited. Figure 1(c) shows the measurement setup for the SDE, where the four-terminal voltage was measured between electrodes 2 and 3, and measurements were carried out using a physical property measurement system (PPMS; Quantum Design). The superconducting trait of FeSe was corroborated by the temperature dependence of the four-terminal resistance, and the midpoint superconducting transition temperature $T_c^{mid}$ was determined to be 13 K (see Fig. 1(d)).

Distinct shifts in the critical current $\Delta I_c = I_{c+} - I_{c-}$, where $I_{c+(-)}$ is the critical current under positive (negative) current application, were detected in FeSe below 10 K without an external magnetic field as manifestation of the field-free SDE in FeSe. Figure 2(a) shows the $I$-$V$ characteristics as a function of temperature of the FeSe device. $\Delta I_c$ increased as the temperature decreased and reached to 92 μA at 4 K. The amplitude of $\Delta I_c$ greatly exceeded the current sweeping step (4 μA). To confirm a response to an external magnetic field, $B_{ext}$, applied perpendicular to the current flow direction, $\Delta I_c$ was measured with respect to the in-plane external magnetic field (see Fig. 2(b)). $\Delta I_c$ is immune to $B_{ext}$ and does not exhibit the odd function behavior with respect to $B_{ext}$ unlike the conventional SDE.

Since FeSe has a symmetric crystal structure and $\Delta I_c$ does not show discernible $B_{ext}$ dependence, the conventional understanding of the SDE, where the breaking of the spatial inversion and time reversal symmetries is simultaneously requested, does not hold in the field-free SDE in FeSe. In particular, the absence of an external magnetic field in the SDE in FeSe is reminiscent of the similar SDE, the field-free Josephson diode effect [16], because the spatial inversion symmetry is broken along the perpendicular to the FeSe; this is analogous to breaking along the stacking plane of $NbSe_2/Nb_3Br_8/NbSe_2$. Meanwhile, the underlying physics of the field-free Josephson diode effect are still elusive, whereas nonreciprocal tunneling with out-of-plane polarization may be a relevant origin. Since a single FeSe layer allows manifestation of the field-free SDE, construction of another model to understand the physics of the field-free SDE in FeSe is required. Indeed, the $\Delta I_c$ is unchanged and exhibits even behavior to the weak perpendicular-to-plane magnetic field within 1 mT, which can be compelling evidence that the field-free SDE in FeSe is not attributable to the Meissner screening enabling the ubiquitous SDE [25] because the ubiquitous SDE is sensitive to the external magnetic field (see also Supplemental Material D [28] for more detail of the control experiments). The fact that the $\Delta I_c$ is immune to the in-plane external magnetic field of up to 9 T allows negating that excess Fe atoms in FeSe play the role of time reversal symmetry breaking as well.

Given that the SDE manifests itself around the critical electric current, where the superconducting to normal (and *vice versa*) transition takes place, and that FeSe exhibits substantial heat response as the large Seebeck coefficient [30], the thermoelectric effect in FeSe can attribute to the findings. The geometry of the FeSe flake is not isotropic but rather triangular-like (see Fig. 1(c)) and the electric current injected from the Au electrode can produce Joule heating due to contact resistance concomitant with the electrode. The heat generated at the contact at the wider side of the FeSe flake dissipates along the in-plane direction more significantly than that at the narrower side since the thermal dissipation is correlated with the thermal conductance, i.e., the width of the FeSe flake. Consequently, the temperature at the narrower side of the flake becomes higher than that at the wider side as shown in Fig. 3(a), and an in-plane temperature gradient $\nabla T$ is generated between electrodes 1 and 4 in the FeSe device (see Supplemental Material E and F [28]). Thermal transport occurs in both normal conductors and superconductors via quasiparticles [29], and the thermal response of FeSe is substantially large (the Seebeck coefficient of FeSe in the normal state at low temperature is approximately −10 μV/K [30]); thus, the in-plane thermal gradient in the FeSe can produce an additional charge current density, $i_{th}$. This results in the addition (cancellation) of the net current density flowing in the FeSe flake when $i_{th}$ is parallel (antiparallel) to $I$ (see also Fig. 3(a)) at a certain temperature and/or in the electric current range, where the superconductivity is close to being broken. When the net current density exceeds the critical current density, $i_c$, the superconductivity of the FeSe breaks and the FeSe transitions to the normal state, which can qualitatively explain the experimental results. Furthermore, we verified that the field-free SDE vanishes in FeSe when the shape of the FeSe flake is not asymmetric, which strongly supports our assertion (see Supplemental Material G [28]). This result also confirms that time reversal symmetry breaking in twin boundaries of FeSe [31] is not the origin of the field-free SDE. Notably, our control experiments using $FeTe_{0.6}Se_{0.4}$ (FTS), which possesses the large Seebeck coefficient in the normal state at low temperature (−10 μV/K [32]) and is the other family of iron-chalcogenide superconductors, also exhibited a similar field-free SDE (see Supplemental Material H [28]).

  To obtain further supporting evidence that the field-free SDE in FeSe is attributed to the thermal gradient-induced phenomenon, we prepared a NbN film (103 nm thick), of which shape is asymmetric (see Fig. 3(b) and [28] for the details of the fabrication). Since the thermoelectric response of NbN is much weaker than that of FeSe (the Seebeck coefficient of NbN in the normal state at low temperature is approximately 0.1 μV/K [33]), we deduced that NbN cannot exhibit a field-free SDE even when shape anisotropy exists. In fact, $\Delta I_c$ of the NbN device is indiscernible as shown in Fig. 3(c), indicating that the field-free SDE is absent in the NbN film with a similar asymmetric structure as the FeSe and FTS investigated in this study, which underscores the validity of our assertion. We note that the device with the NbN channel that is the complete mimic of the FeSe channel in size, structure and geometry exhibited suppression of the SDE (see Supplemental Material J [28]), which is strong supporting evidence that the field-free SDE of FeSe is not attributed to Meissner screening and/or asymmetric vortex barriers [25,34,35] that are sensitive to asymmetric or serrated edge structures of superconductors. Furthermore, the FeSe devices with a sharp and a serrated edges exhibit almost the same field-free SDE, which is additional evidence that the thermoelectric response in FeSe, not the edge structure of it, plays a pivotal role in manifestation of SDE (see Supplemental Material K [28]). Here, we note two issues: First, albeit the heat flow can break time reversal symmetry as well, this type of the breaking is effective time-reversal symmetry breaking under non-equilibrium unlike the breaking under equilibrium, such as an external magnetic field, which is notable novelty in the physical finding. Second, the $\Delta I_c$ of the FeSe device observed in the measurements exhibits discrepancy of more than two orders of magnitude to the calculated $\Delta I_c$ using the Seebeck coefficient and estimated $\nabla T$ (see Supplemental Material F and L for the detail [28]), which is a significant open question for future studies. Meanwhile, our assertion that the combination of thermoelectric response and temperature gradient in FeSe gives rise to the field-free SDE found in this study is corroborated by the experiment that is described in the next paragraph.

The compelling evidence to support our claim is the successful sign reversal of the polarity of the SDE in FeSe by the external heat flow. Figure 4(a) shows the optical image of the device and measurement setup. An individual heating electrode made of Al wire is equipped close to the FeSe device, which enables us to control the thermal gradient of the FeSe by the heating effect at the Al heating wire. As shown in Figs. 4(b) and 4(c), the polarity of the SDE is positive ($I_{c+}$ is greater than $I_{c-}$), which is attributed to the fact the temperature at the contact 2 side ($T_2$) is higher than that at the contact 1 side ($T_1$) due to the smaller contact area of the electrode 2.

Meanwhile, when the heat creation at the Al heating wire takes place, the heat gradient in the FeSe is reversed thanks to the heat flow from the Al heating wire to the FeSe (see Fig. 4(d)). In consequence, the sign reversal of the SDE polarity can be expected when the heating at the Al wire is on. As expected, the polarity of the SDE is controllable and reversed by the heat generation by injecting the electric current of 30 mA to the Al wire as shown in Figs. 4(c) and 4(e). Furthermore, the sign reversal is collaborated at each measuring temperature (from 4 to 8 K, see Figs. 4(f)-(h)). To note is that the decrease of the $\Delta I_c$ after the sign reversal (for example, see the data of the $I_{heat}$ of 30 and 35 mA in Fig. 4(f)) is attributed to the fact that the temperature of the FeSe goes up by the heating and the superconductivity is approaching to be broken. This result unequivocally shows that the heat effect governs the SDE in FeSe. In consequence, all field-free SDE results obtained in this study using materials with various thermal responses unequivocally show that the origin of the field-free SDE in FeSe is the thermal gradient due to asymmetry in the shape of the FeSe flake, i.e., the intrinsic effect unlike extrinsic one [25].

In summary, we found a field-free SDE in a layered superconductor FeSe film, the origin of which was determined to the in-plane temperature gradient due to Joule heating generated at the contact resistance and asymmetric geometry of the FeSe device, i.e., we found a novel intrinsic mechanism enabling the SDE. Despite the simple device structure composed of only FeSe flake and nonmagnetic electrodes, nonreciprocal critical current between the superconducting and normal states was successfully observed. Since the SDE induced by heat dissipation was generated simply by shaping a material with a large thermoelectric coefficient into an asymmetric shape, our findings can open a novel pathway for achieving energy-efficient logic devices utilizing the SDE. Furthermore, our finding shed light on significance of effective time-reversal symmetry breaking under non-equilibrium unlike the breaking under equilibrium, such as an external magnetic field, and also requests revisiting previous experimental setups in the field-free SDE, where asymmetric structures in superconducting channels and/or materials exhibiting large thermal response are exploited, in view of thermal dissipation.

**Acknowledgements**

This work is partly supported by JSPS Grant-in-Aid for Challenging Research (Pioneering) (23K17353,

21K18145 and 19H05522), Grant-in-Aid for Scientific Research (B) (22H01181), Grant-in-Aid for Scientific Research (S) (22H04933), Grant-in-Aid for Early-Career Scientists (21K13880), and Grant-in-Aid for Transformative Research Areas (24H01662). U.N., M.A., R.O, Y.A. and M.S. would like to thank Dr. E. Tamura for his fruitful discussion.

**Data availability**

The data that support the findings of this study are available from the corresponding author upon reasonable request.

**FIGURES and FIGURE CAPTIONS**

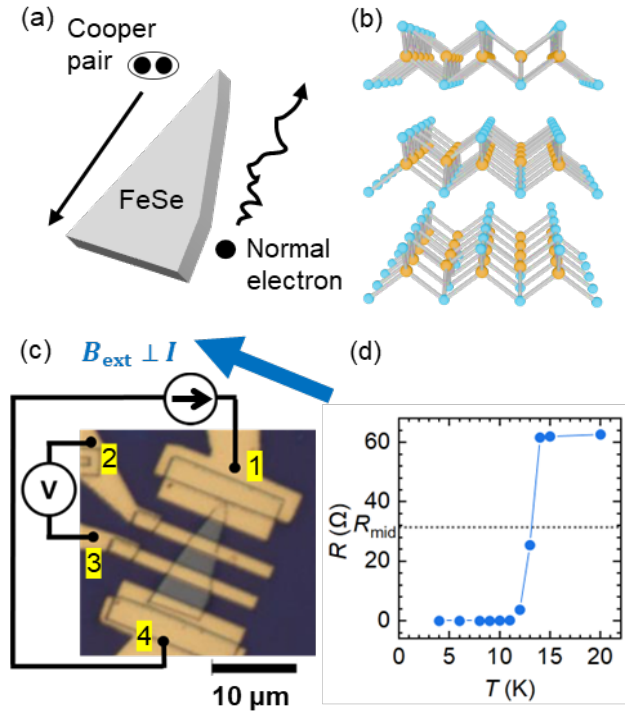

**FIG. 1 (a)** Schematic of the SDE in FeSe, where the Cooper pair and the normal charge current flow in opposite directions. **(b)** Crystal structure of FeSe. **(c)** Measurement setup and the fabricated device image used in our experiment. The width of electrodes 2 and 3 was 1.5 μm. The absolute value of the applied current was increased from 0 μA to a maximum value in steps of 4 μA. **(d)** Temperature dependence of the four-terminal resistance of FeSe in the device shown in **(c)**. $R_{mid}$ denotes the half resistance of the normal conduction resistance.

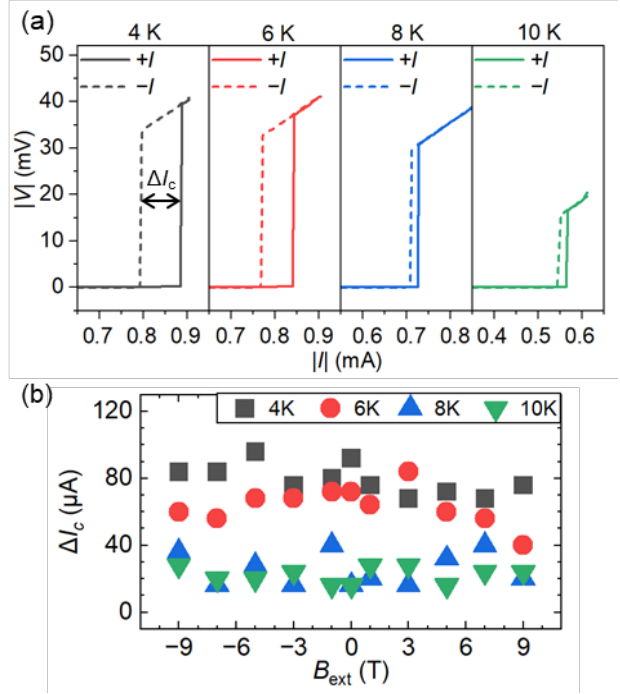

**FIG. 2 (a)** *I-V* characteristics of the device at from 4 to 10 K under positive (the solid lines) and negative (the dashed lines) electric currents. The black, red, blue, and green lines show the *I-V* curves at 4, 6, 8, and 10 K, respectively. An external magnetic field was not applied. **(b)** External magnetic field dependence of $\Delta I_c$ at temperatures ranging from 4 to 10 K. The black squares, red circles, blue up-pointing triangles, and green down-pointing triangles show the values of $\Delta I_c$ at 4, 6, 8, and 10 K, respectively.

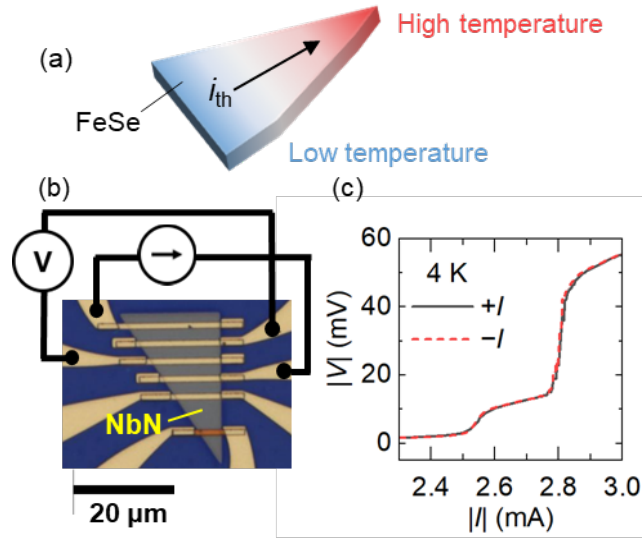

**FIG. 3 (a)** Schematic of a thermal gradient in FeSe, resulting in an additional current flow. $i_{th}$ denotes the thermoelectric current generated by the thermal gradient. **(b)** Measurement setup and the fabricated device for the control experiment using asymmetric NbN. An external magnetic field is not applied. **(c)** *I-V* characteristics of the NbN device at 4 K under positive (black solid line) and negative (red dashed line) electric current injection. The absolute value of the applied current was increased from 0 mA in 0.004 mA steps. The field-free SDE is not observed.

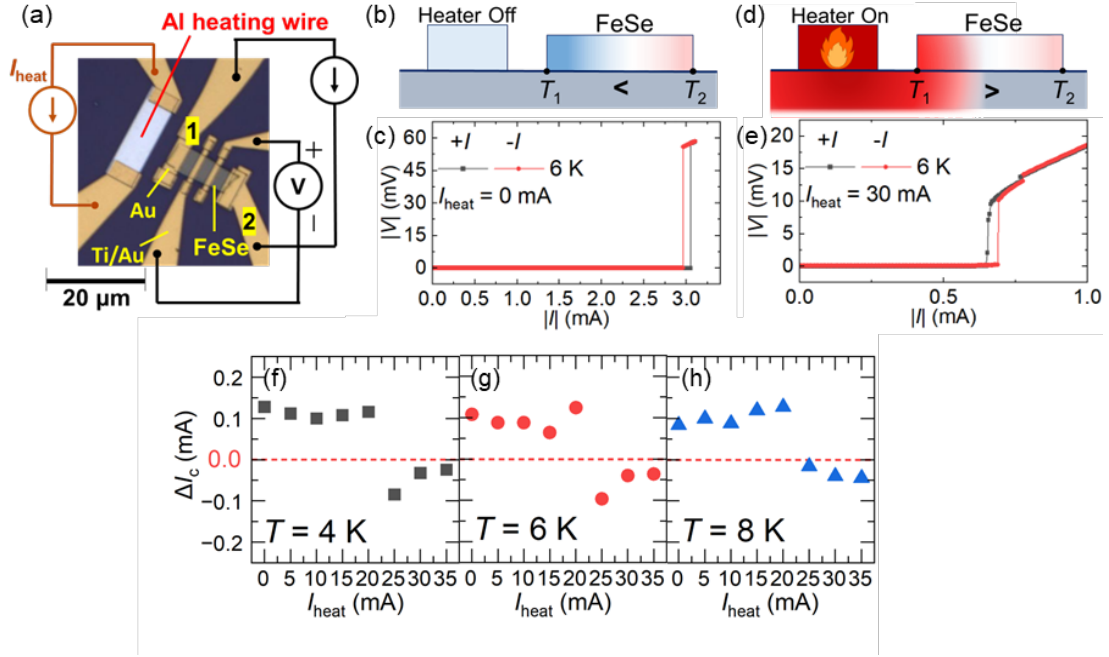

**FIG. 4 (a)** Optical image of the device and measuring setups. **(b)** Schematic image of the heart gradient in FeSe when the heater is off. **(c)** The SDE of the FeSe at 6 K without heating effect at the Al heating wire. The $I_{c+}$ is greater than the $I_{c-}$, and the polarity of the SDE is positive since the temperature at contact 2 is higher than that at contact 1 ($T_1 < T_2$). **(d)** Schematic image of the heart gradient in FeSe when the heater is on. **(e)** The SDE of the FeSe at 6 K with heating effect at the Al heating wire. The injected electric current to the Al heating wire was set to be 30 mA. The $I_{c+}$ becomes smaller than the $I_{c-}$ by the reversal of the heat gradient to $T_1 > T_2$ in the FeSe by the heating effect at the Al heating wire. The polarity of the SDE is hence changed to be negative. **(f)-(h)** The polarity changes in the SDE of the FeSe by the heating effect at the Al heating wire at 4, 6, and 8 K.